\documentclass[letterpaper,10pt,onecolumn]{article}
\usepackage{tabularx} 
\usepackage{amsmath}  
\usepackage{graphicx} 
\usepackage[margin=1in]{geometry} 
\usepackage[final]{hyperref} 
\hypersetup{
	colorlinks=true,       
	linkcolor=blue,        
	citecolor=blue,        
	filecolor=magenta,     
	urlcolor=blue         
}
\usepackage{graphicx}

\usepackage{subcaption}
\usepackage{blindtext}
\usepackage[numbers,sort]{natbib}
\usepackage[utf8]{inputenc} 
\usepackage[T1]{fontenc}    
\usepackage{url}            
\usepackage{booktabs}       
\usepackage{amsfonts}       
\usepackage{nicefrac}       
\usepackage{microtype}      
\usepackage{xcolor}         
\usepackage{mathrsfs}

\usepackage{dsfont}
\usepackage{amssymb}  
\usepackage{makecell}
\usepackage{amsmath}
\usepackage{multirow}
\usepackage{booktabs}
\usepackage{float}
\usepackage{xurl}          
\hypersetup{breaklinks=true}
\usepackage{algpseudocode}
\usepackage{amsthm}
\theoremstyle{definition}

\usepackage{xspace}
\usepackage{longtable}
\usepackage{wrapfig}
\usepackage{enumitem}
\usepackage{etoc}

\usepackage{array}
\setlist{leftmargin=5mm}
\usepackage[ruled,vlined]{algorithm2e}
\usepackage[export]{adjustbox}

\usepackage{todonotes}
\usepackage{xcolor}

\begin{document}

\author{
  \textbf{Jio Oh\textsuperscript{1}\textsuperscript{2}\textsuperscript{4}\footnotemark[1]},
  \textbf{Steven Euijong Whang\textsuperscript{1}},
  \textbf{James Evans\textsuperscript{3}},
  \textbf{Jindong Wang\textsuperscript{4}}\footnotemark[2]  \\\\
    \textsuperscript{1}KAIST,
  \textsuperscript{2}Microsoft Research Asia,
  \textsuperscript{3}University of Chicago,
  \textsuperscript{4}William \& Mary
}

\date{}
\title{Classroom AI: Large Language Models as Grade-Specific Teachers}
\maketitle
\footnotetext[1]{Work done during internship in Microsoft Research Asia and William \& Mary \texttt{<harryoh99@kaist.ac.kr>}}
\footnotetext[2]{Corresponding author: Jindong Wang \texttt{<jdw@wm.edu>}.}
\footnotetext[0]{This paper was published in npj Artificial Intelligence (2026).}

\begin{abstract}
Large Language Models (LLMs) offer a promising solution to complement traditional teaching and address global teacher shortages that affect hundreds of millions of children, but they fail to provide grade-appropriate responses for students at different educational levels. We introduce a framework for finetuning LLMs to generate age-appropriate educational content across six grade levels, from lower elementary to adult education. Our framework successfully adapts explanations to match students' comprehension capacities without sacrificing factual correctness. This approach integrates seven established readability metrics through a clustering method and builds a comprehensive dataset for grade-specific content generation. Evaluations across multiple datasets with 208 human participants demonstrate substantial improvements in grade-level alignment, achieving a 35.64 percentage point increase compared to prompt-based methods while maintaining response accuracy. AI-assisted learning tailored to different grade levels has the potential to advance educational engagement and equity. 
\end{abstract}

\section*{Introduction}
\label{Introduction}
\begin{figure*}[t]
    \centering
    \includegraphics[width=\linewidth]{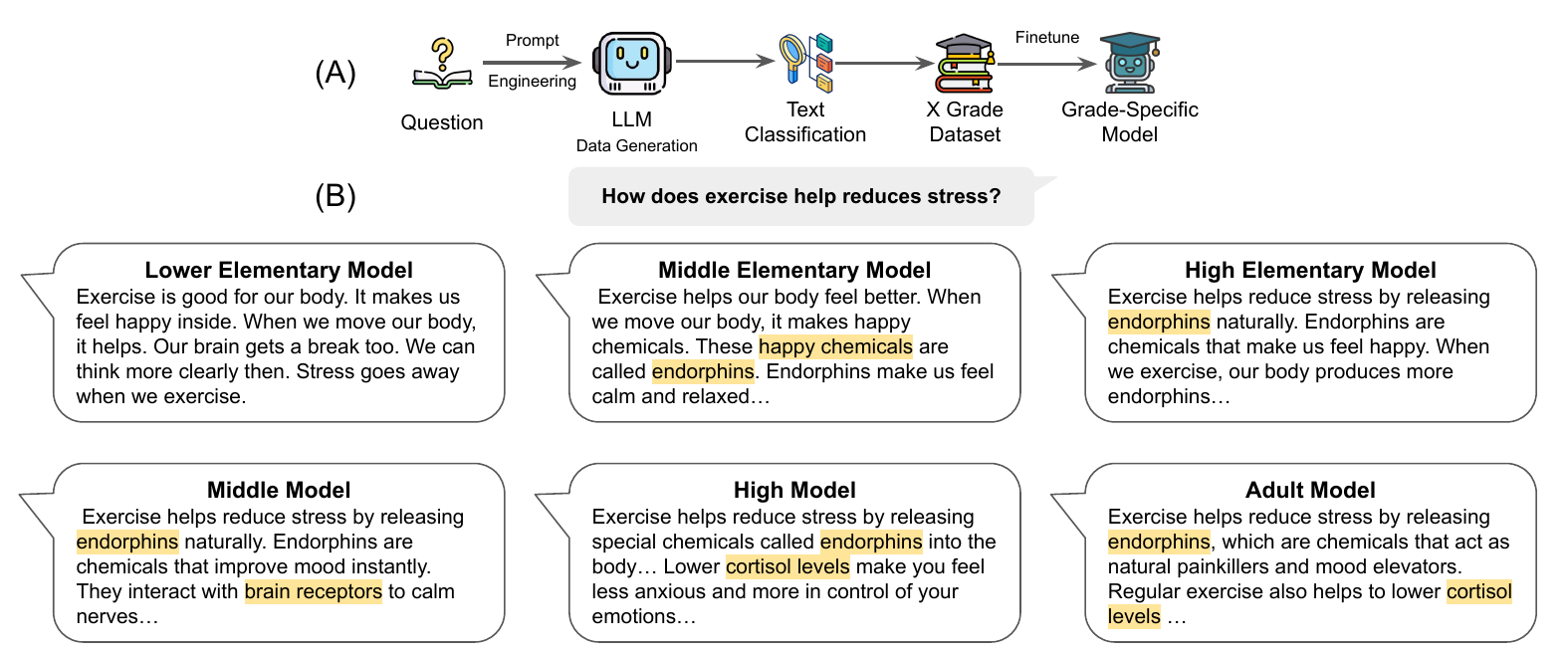}
    \caption{(A) Overall framework of our approach. (B) Exemplar responses across different models.} 
    \vspace{-0.3cm}
    \label{fig:framework}
\end{figure*}
Large Language Models (LLMs) have expanded beyond traditional natural language processing into diverse domains ranging from finance~\cite{survey2023finance,lee2024survey} and healthcare~\cite{nazi2024largelanguagemodelshealthcare,tam2024frameworkhumanevaluationlarge} to education~\cite{huang2024socialsciencemeetsllms,wang2024largelanguagemodelseducation,gan2023largelanguagemodelseducation}. With models like GPT~\cite{openai2024gpt4technicalreport}, Gemini~\cite{gemini}, and LLaMA~\cite{grattafiori2024llama3herdmodels} now accessible worldwide, LLMs transform education by assisting teachers and students with question-solving, confusion clarification, material creation, and content personalization~\cite{gan2023largelanguagemodelseducation, wang2024largelanguagemodelseducation}.

Teacher shortages present a large and growing global challenge. UNESCO estimates 44 million additional teachers are needed to achieve universal primary and secondary education by 2030~\cite{unesco2022outofschool}, while 35\% of U.S. public schools report at least one teaching vacancy~\cite{nces2024pressrelease}. Worldwide, 244 million children lack access to school~\cite{unesco2022outofschool}, with shortages most acute in rural and high-poverty areas~\cite{garcia2019teacher}. In Pakistan, 44\% of students drop out between ages 5-16~\cite{shah2018pakistan, khan2021child}, with 18.6\% leaving before completing primary education. Limited access to qualified teachers exacerbates educational disparities~\cite{garcia2019teacher,khan2021child}, with over 70\% of teachers in sub-Saharan Africa classified as inadequately trained~\cite{sage2024worldteachersday}.

With 67.9\% of the global population connected to the internet, LLM-based educational tools could benefit over 100 million children currently without school access and provide enhanced assistance to countless others. LLMs can deliver consistent explanations and personalized assistance regardless of geographic location, potentially increasing learning engagement and reducing educational inequity worldwide.

Effective teachers require both subject knowledge and pedagogical skills tailored to different grade levels. Despite their capabilities, LLMs struggle to provide grade-appropriate answers~\cite{rooein2023knowaudiencellmsadapt}. Even with explicit prompts like ``Answer for 3rd graders'', LLMs generate responses that systematically exceed the target grade's comprehension level. Existing works mostly focus on prompt-based evaluations that fail to achieve a satisfactory level due to the lack of comprehensive evaluation criteria and appropriately curated training corpora~\cite{imperial-tayyar-madabushi-2023-flesch,gobara2024llmsimplicitlydeterminesuitable, Hsu2024FreetextRG,10.1145/3627508.3638345}. To serve as effective educational tools, LLMs must produce content students can understand and directly engage with.

Previous work on finetuning LLMs for specific reading levels focused primarily on summarization or paraphrasing~\cite{malik-etal-2024-tarzan,tran2024readctrlpersonalizingtextgeneration, ribeiro2023generating}. However, real classroom settings involve open-ended questions without source texts to summarize or paraphrase. For example, when a student asks ``What is gravity?'' and a teacher needs LLM assistance, summarization approaches fail. This becomes critical in AI tutoring scenarios where students interact directly with LLMs. Even high-quality LLM responses provide no benefit if students cannot comprehend them.

We propose a framework for grade-level targeted finetuning of LLMs that handles open-ended educational queries across various subjects. Our approach enables grade-appropriate content generation for natural questions across six educational levels: lower elementary (grades 1-2), middle elementary (grades 3-4), upper elementary (grades 5-6), middle school (grades 7-9), high school (grades 10-12), and college/adult (grade 13+). This granular classification reflects research in educational psychology showing that reading and comprehension skills develop rapidly during early education~\cite{chall1983stages, national2000teaching}. This classification can provide a more accurate measure of grade-level suitability, enabling our evaluation framework to detect subtle yet significant shifts in linguistic complexity that might be overlooked. Moreover, our approach advocates for models that behave appropriately given teaching context~\cite{leibo2024theory}. 

To assess text complexity, we integrate seven established readability metrics: Flesch Reading Ease~\cite{flesch1948new}, Flesch-Kincaid Grade Level~\cite{kincaid1975derivation}, the Coleman-Liau Index~\cite{coleman1975computer}, Linsear Write~\cite{o1966gobbledygook}, the Gunning Fog Index~\cite{gunning1952technique}, Dale-Chall~\cite{chall1995readability}, and the Spache Readability Formula~\cite{spache1953new}. Each metric captures different aspects of readability, and we group them based on their underlying characteristics to create a more reliable integrated measure (see \nameref{subsec:Readability Metrics Explaination} for further details).

To address the challenge of limited training data for grade-specific content, we generate data using LLMs, a technique increasingly used for data collection and generation~\cite{li2023syntheticdatagenerationlarge,long2024llmsdrivensyntheticdatageneration}. LLMs are known to generate high-quality text data that aligns with user instructions, thereby improving model performance when finetuned on such data~\cite{peng2023instructiontuninggpt4,chen2024selfplayfinetuningconvertsweak, alpaca}. Notably, Orca-math~\cite{mitra2024orca} presents a math dataset with GPT by guiding the model to adopt a teacher–student paradigm, which highlights the potential of instruction-aligned synthetic data to enhance educational resources. We categorize 54 subjects across 8 fields to create diverse questions, then use state-of-the-art LLMs (GPT~\cite{openai2024gpt4technicalreport}, Gemini~\cite{gemini}, LLaMA~\cite{grattafiori2024llama3herdmodels}, and Mixtral~\cite{jiang2024mixtralexperts}) to generate questions answerable across all grade levels (see Supplementary Figure 1, for details on subjects). We craft tailored prompts for LLaMA3.1:70B to produce outputs for each grade level, then classify generated text using our integrated metrics algorithm (see \nameref{subsec:answerprompt} for prompt details).

Our experiments demonstrate that this approach significantly improves grade-level alignment compared to prompt-based methods while maintaining response accuracy. Human studies with 208 participants confirm that our framework aligns with human perceptions of grade-appropriate content.

Our contributions include:

\begin{itemize}
    \item A framework for developing grade-specific LLMs to enhance educational equity and deliver social benefits globally.
    \item Extensive evaluation with 208 participants validating our framework's alignment with human perceptions of difficulty, showing that finetuned models can explain complex concepts at targeted grade levels.
    \item A model-agnostic dataset for finetuning LLMs in educational contexts, integrating multiple educational metrics with grade-appropriate responses for open-ended questions.
\end{itemize}

\section*{Results}
\begin{figure*}[htbp]
    \centering
    \includegraphics[width=\linewidth]{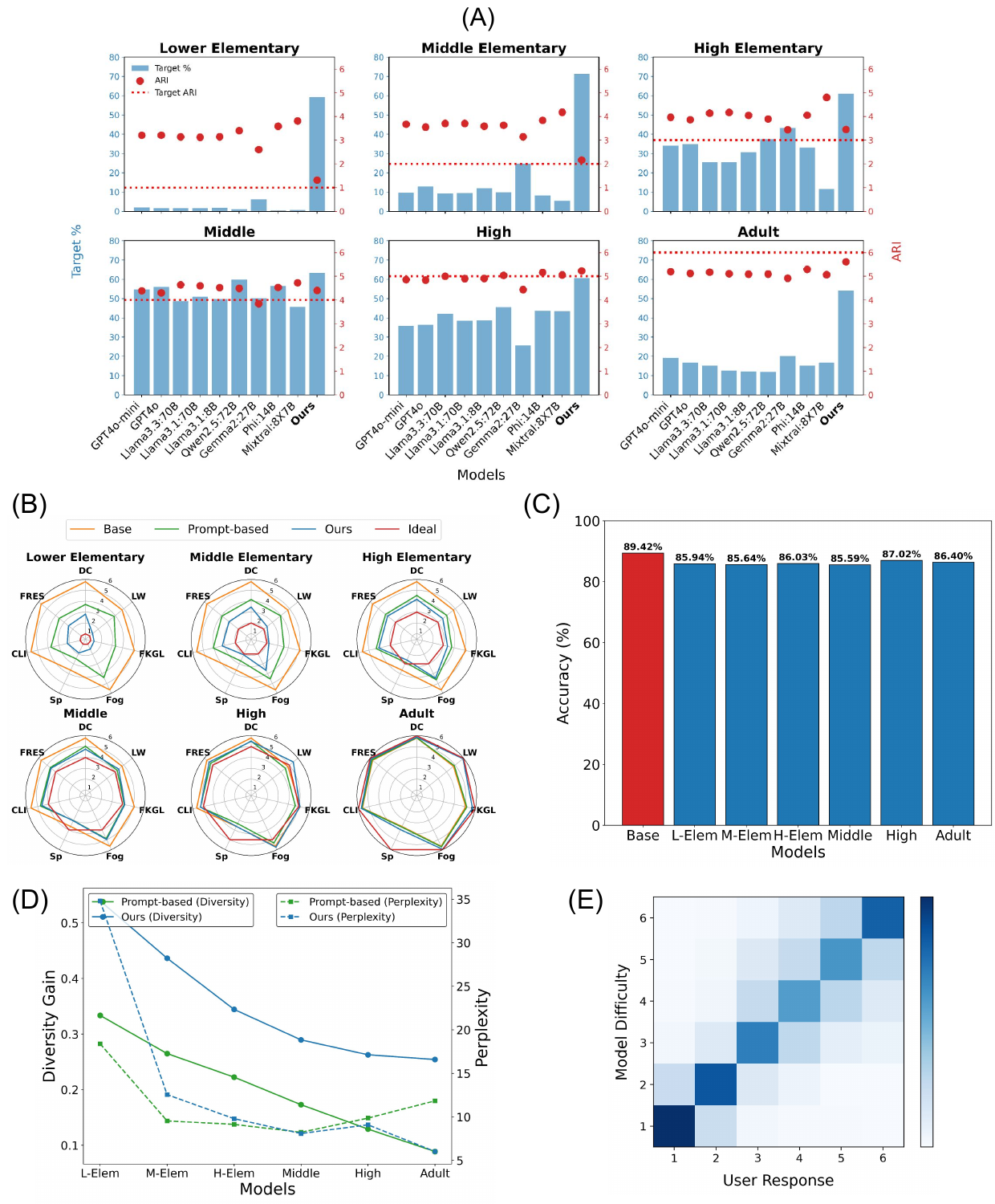}
    \caption{Results across evaluation criteria: (A) compatibility through integrated measure and ARI; (B) compatibility for each metric; (C) accuracy; (D) diversity gain and perplexity; and (E) survey results for type 1 questions.} 
    \vspace{-0.5cm}
    \label{fig:whole_res}
\end{figure*}

We evaluate our method on two main dimensions: \textbf{compatibility} and \textbf{accuracy}. \textbf{Compatibility} measures whether finetuned models' outputs align with target grade students' comprehension capability. We assess compatibility using: (1) an integrated measure using seven readability metrics (see Supplementary  Section B, for definitions of the metrics), (2) individual evaluations for each metric, and an (3) Automated Readability Index (ARI)~\cite{smith1967automated} as a held-out metric to test generalizability. \textbf{Accuracy} measures whether the model maintains its ability to generate correct and relevant responses.

We also measure perplexity and diversity gain \cite{bilmes2022submodularitymachinelearningartificial}, which reflect linguistic variety relative to the training corpora and the base model. Finally, we conduct surveys with 208 human participants and GPT4o to validate our framework's alignment with human perceptions.

\subsection*{Compatibility}
\label{subsec: res compat}

We test the finetuned models' compatibility on all four datasets. For $\mathcal{D}_{GPT}$, $\mathcal{D}_{ELI}$, and $\mathcal{D}_{NQ}$, we use all sampled questions. For $\mathcal{D}_{SQ2}$, we split questions based on their designated grade levels and analyze results accordingly. For example, when targeting lower elementary level, we focus exclusively on grades 1-2 questions.

As shown in Figure~\ref{fig:whole_res}(A), our approach significantly increases target success rates for each grade level compared to prompt-based approaches, with an average improvement of 35.64 percentage points over the prompt-based baseline. The blue bars represent the success rate of each model in producing outputs at the intended grade level (higher is better), while red dots show corresponding ARI values discretized into six difficulty levels. Dots closer to the red dotted line indicate stronger alignment with ARI. Similar improvements appear for the held-out ARI metric, with our approach best aligning with intended grade levels. Detailed output grade-level distribution is shown in Supplementary Section C.3 and C.6.

Figure~\ref{fig:whole_res}(B) shows that our approach successfully shifts all seven educational metrics toward their optimal values (1 for lower elementary through 6 for adult) compared to the base model or prompt-based approaches. (Note that DC, LW, FKGL, Fog, Sp, CLI, and FRES are measures for Dale-Chall, Linsear Write, Flesch-Kincaid Grade Level, Gunning Fox Index, Spache Readability Formula, Coleman-Liau Index, and Flesch Reading Ease respectively.) This improvement stands out for elementary school grade levels, which previous research identified as most challenging to target~\cite{rooein2023knowaudiencellmsadapt}.

\subsection*{Accuracy}
\label{subsec: res_acc}

We test the finetuned models' accuracy on $\mathcal{D}_{SQ}$, a multiple-choice dataset aligned with educational contexts. Figure~\ref{fig:whole_res}(C) shows that finetuned models achieve performance comparable to the base model. While finetuning typically causes some accuracy reduction~\cite{wang2024twostagellmfinetuningspecialization, dodge2020fine, luo2025empiricalstudycatastrophicforgetting}, our results show minimal performance degradation.

\subsection*{Perplexity and Diversity}
\label{subsec: res_div}

We measure output unexpectedness through perplexity and diversity gain (see Supplementary Section B.2 and B.3, for formulas). Figure~\ref{fig:whole_res}(D) shows that lower-grade models exhibit higher values for both metrics, suggesting that text comprehensible for lower grades appears less frequently in training corpora, explaining why traditional approaches struggle with these levels. Lower-grade models convey difficult concepts using simpler language, resulting in higher diversity compared to the more direct language of existing models.

\subsection*{Survey}
\label{subsec: res_survey}

\begin{figure*}[htbp]
    \centering
    \includegraphics[width=0.7\linewidth]{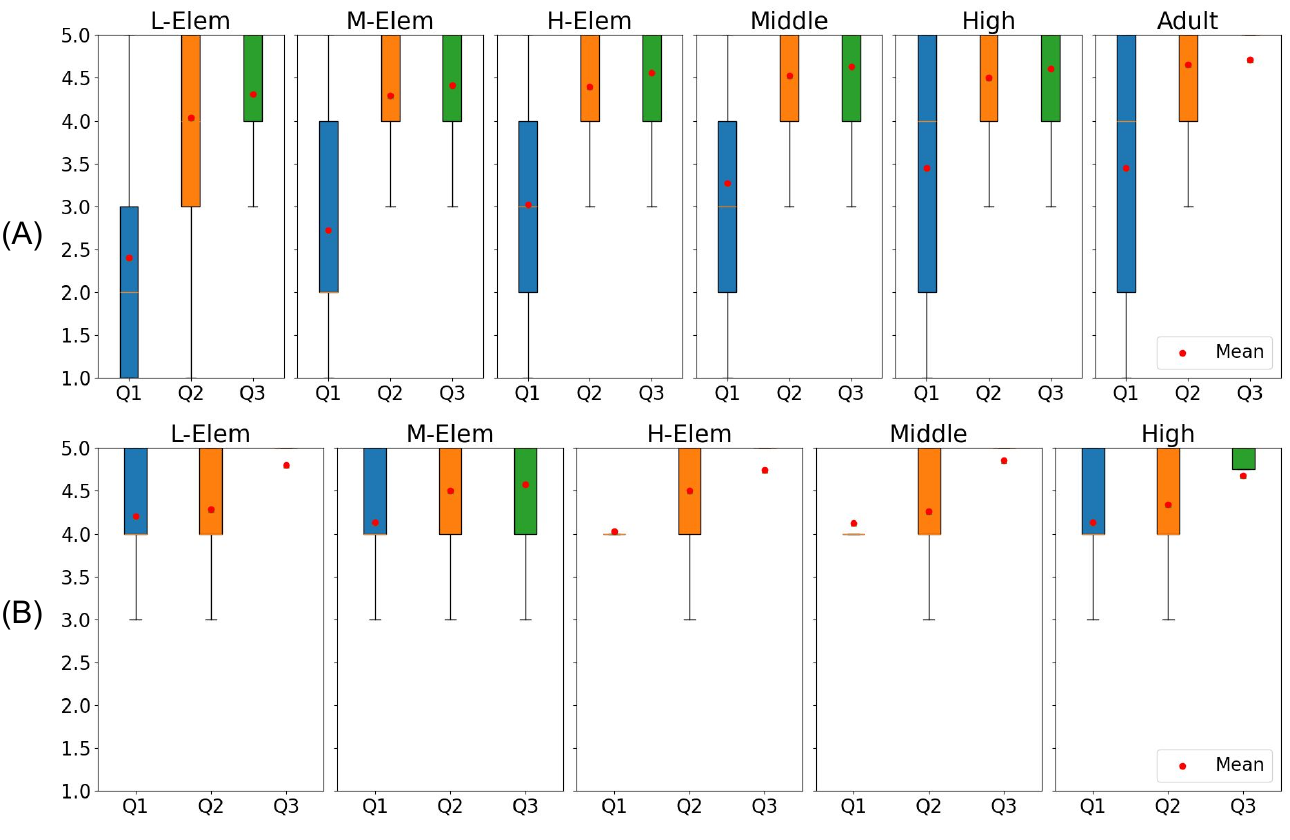}
    \caption{Survey results on Type 2 questions: (A) $\mathcal{D}_{NQ}$ and (B) $\mathcal{D}_{SQ2}$. Box plots show Q1 (question difficulty), Q2 (answer comprehensibility), and Q3 (model accuracy) across grade levels, with mean values as red dots on a five-point scale. Higher Q1 and Q2 results indicate lower question and answer difficulty; higher Q3 results indicate stronger accuracy. While $\mathcal{D}_{NQ}$ questions appear difficult for lower grades (low Q1 in A), answers remain comprehensible (high Q2 in A). Answer comprehensibility increases (higher Q2 in B vs. A) for grade-specific questions in $\mathcal{D}_{SQ2}$.} 
    \vspace{-0.3cm}
    \label{fig:box_plots}
\end{figure*}
We conduct two surveys using different datasets. Survey 1 uses questions from $\mathcal{D}_{NQ}$ with 108 participants, while Survey 2 uses questions from $\mathcal{D}_{SQ2}$ with 120 participants. All participants are English-speaking and have completed high school, with most being undergraduate or graduate students who understand the relative difficulty across grade levels. We use two question types:

\begin{itemize}
    \item Type 1: Each question includes six answers from different finetuned models ranging from lower elementary to adult. Participants assign each answer to a unique grade level, effectively ranking them.
    \item Type 2: Each question includes one answer from a finetuned model. Participants answer three five-point scale questions about question difficulty (Q1), answer comprehensibility (Q2), and answer accuracy (Q3).
\end{itemize}

For Type 1 questions, we measure the association between human-perceived difficulty rankings and model outputs using Kendall's $\tau$ coefficient. The high coefficient of 0.76 across 108 participants demonstrates strong agreement between intended and perceived difficulty levels. We also compute L1 distances between participant rankings and ground truth ordering. For example, given six outputs sorted by ascending grade levels, the ground-truth ranking would be [1,2,3,4,5,6]. If a participant ranks them as [6,5,4,3,2,1], the L1 distances would be [5,3,1,1,3,5]. The L1 distances between rankings, [0.293, 0.398, 0.659, 0.693, 0.676, 0.578], all fall below 1. Combined with the high Kendall's $\tau$ coefficient of 0.76, these results confirm our approach successfully generates grade-appropriate text aligned with human perception. Figure~\ref{fig:whole_res}(E) shows dark cells along the diagonal, indicating strong alignment between model outputs and human perceptions.

For Type 2 questions, Figure~\ref{fig:box_plots} shows user scores on a five-point scale for Q1, Q2, and Q3 across two surveys. For $\mathcal{D}_{NQ}$, despite relatively difficult questions (low Q1 scores), models generate outputs comprehensible for each grade level. In post-survey feedback, participants noted that lower-grade models effectively explain concepts beyond their grade level using shorter, simpler sentences.

Regarding relatively lower Q2 scores (answer comprehensibility) for lower grades, participants attributed this to topic complexity rather than explanation difficulty. One noted, ``Despite how clearly the model explains the concept of LLC sublayer in an operating system, lower elementary students will struggle with the concept itself, hence I give a 1 for Q2.'' This observation is supported by higher Q2 scores in the second survey ($\mathcal{D}_{SQ2}$), where questions were designed for specific grade levels. High Q3 scores across all grade levels confirm our approach maintains factual correctness while adapting explanations.

We also evaluate surveys using GPT4o, following recent trends in using LLMs for human value alignment~\cite{zheng2023judgingllmasajudgemtbenchchatbot,li-etal-2024-leveraging-large}. Results from GPT4o align with human evaluations, further validating our approach (see Supplementary Section C.7, for more details).

\section*{Discussion}
\label{subsec:world_view}

\begin{figure*}[h!]
    \centering
    \includegraphics[width=0.9\linewidth]{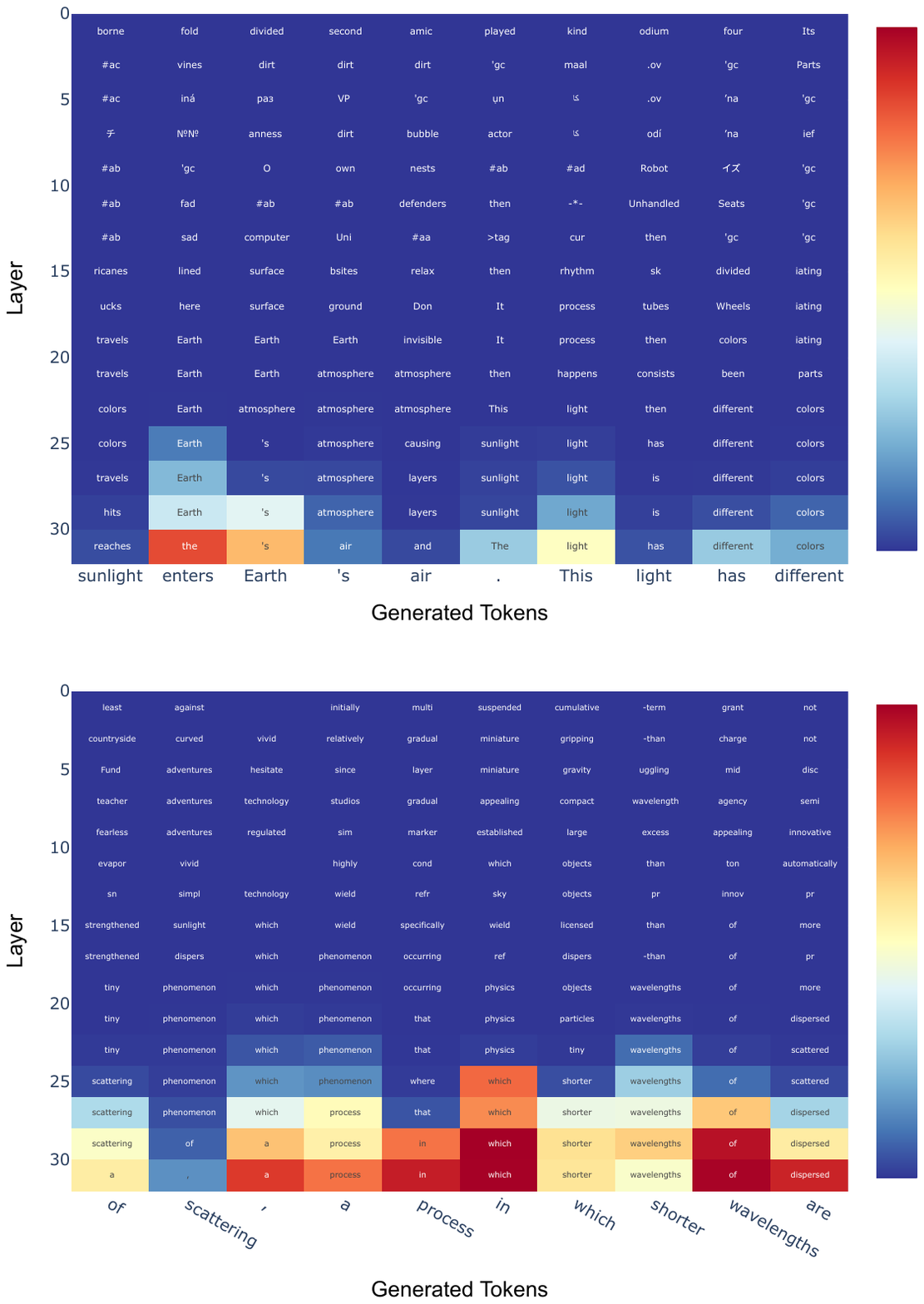}
    \caption{Logit-lens visualization for the lower-elementary (top) and adult (bottom) models of LLaMA3.1:8B on the prompt, ``Why is the sky blue? The sky is blue because''. The bottom row for each figure shows the final output tokens, and each row above represents the top prediction at each transformer layer. Warmer colors (\emph{e.g.}, red) indicate higher confidence.} 
    \vspace{-0.3cm}
    \label{fig:logitlens}
\end{figure*}

As students' perspectives evolve with age, we investigate whether finetuned models develop distinct worldviews by examining model layers and outputs. We finetune LLaMA3.1:8B and compare results between lower elementary and adult models. Using logit lens~\cite{belrose2023eliciting} to analyze internal model layers, we observe that lower-grade models formulate ideas more directly and succinctly, while higher-grade models favor in-depth explanations as shown in Figure~\ref{fig:logitlens}. This pattern mirrors human cognitive development, emphasizing clarity for younger students while preserving detail for advanced audiences. Plus, for lower-grade models, certain complex words (\emph{e.g.}, ``atmosphere'') are replaced by simpler synonyms (\emph{e.g.}, ``air'') and higher grade models internally show relatively complicated words such as ``wavelength'' or ``dispersed''.

Moreover, analysis of vocabulary and sentence structure reveals that lower-grade models use simpler words and shorter sentences, while higher-grade models employ specialized terminology with longer explanations. These findings indicate our finetuning approach influences not only readability, but also the way models think and communicate, aligning each model with its target audience's comprehension needs. Detailed visualizations appear in Supplementary Section D.

Our research addresses the critical global teacher shortage that impacts millions of children, aiming to improve educational equity. Our approach introduces a novel framework for training grade-specific LLMs to deliver age-appropriate educational content. These tools can supplement traditional teaching, providing personalized support to students of different grades and potentially increasing learning engagement worldwide. We believe that our work can contribute to a future where LLM-assisted learning can help mitigate educational disparities and create broader social benefits. The authors are responsible for all analyses and the final manuscript content.

While our approach successfully tailors textual complexity, it does not fully address conceptual difficulty. For example, even when written at a lower reading level, concepts like organizational culture may remain difficult for young students to comprehend based on their limited experience with the organizational world. Future work could incorporate domain-specific knowledge graphs or concept taxonomies to provide step-by-step explanations of challenging ideas. Combining readability metrics with conceptual difficulty frameworks would create truly adaptive LLMs that match both linguistic capacity and conceptual background.

\section*{Methods}
\label{sec:Methods}
We construct a grade-aligned question–answering dataset by generating diverse questions  across eight educational fields (\nameref{subsubsec:q_generate}) and prompting LLMs to produce answers with different readability levels (\nameref{subsec:answerprompt}). We then assign grade-levels to the generated data using a novel integrated metric based on seven established readability formulas (\nameref{subsec:Readability Metrics Explaination}) and finally fine-tune six grade-specific models to produce grade-appropriate responses (\nameref{subsec:model ft}) and test the performance of the models across different datasets (\nameref{subsec:datasets}). An overview of the full framework is illustrated in Figure~\ref{fig:framework}.

\subsection*{Question Generation}
\label{subsubsec:q_generate}
We define eight educational fields based on K-12 curriculum frameworks: art, artificial intelligence, health education, literature, music, physical education, science, and social science. Each field contains five to eight subjects (see Supplementary Figure 1). To create a comprehensive question set, we prompt ChatGPT to generate sample questions answerable across all grades for each subject. Using these generated questions as few-shot demonstrations~\cite{openai2024gpt4technicalreport}, we employ LLMs including Gemini~\cite{gemini}, GPT~\cite{NEURIPS2020_1457c0d6}, and LLaMA~\cite{grattafiori2024llama3herdmodels} to generate approximately 550 questions per subject.

\subsection*{Answer Generation}
\label{subsec:answerprompt}

We design prompts to align with readability metrics and prior works~\cite{rooein2023knowaudiencellmsadapt,Hsu2024FreetextRG} by varying word difficulty, sentence length, and target audience. The prompt template follows: \textit{``Please provide the explanation in plain text with no bullet points using <very easy, fairly easy, fairly difficult> words that <elementary school 1st grade, elementary school 3rd grade, elementary school 5th grade, middle school 7th grade, high school 10th grade, or college> students will know. Answer in detail with at a maximum of <4, 5, 6, 7, 8, 10, 12, 15, or 20> words per sentence.''}. The output distribution for the corresponding prompts is shown in Supplementary Figure 3.

We vary grade level and maximum sentence length in the input prompts, creating 54 (6×9) distinct question-answer pairs for each question. Word difficulty matches grade level based on the Flesch Reading Ease Metric~\cite{flesch1948new}. Using our integrated metric, we classify Q\&A pairs into six grade levels. The distribution appears in Supplementary Table 1. 
\subsection*{Readability Metrics Integration}
\label{subsec:Readability Metrics Explaination}

We integrate seven readability metrics: Flesch Reading Ease (FRES)~\cite{flesch1948new}, Flesch-Kincaid Grade Level(FKGL)~\cite{kincaid1975derivation}, the Coleman-Liau Index(CLI)~\cite{coleman1975computer}, Linsear Write(LW)~\cite{o1966gobbledygook}, the Gunning Fog Index(Fog)~\cite{gunning1952technique}, Dale-Chall(DC)~\cite{chall1995readability}, and the Spache Readability Formula(Sp)~\cite{spache1953new}.

Each metric captures distinct aspects of linguistic complexity through word difficulty, sentence length, and syntactic structure. Because their calculations differ substantially, we categorize them into three groups based on shared characteristics to create an integrated evaluation process:

\[
\begin{aligned}
    G_1 &= \text{Metrics utilizing predefined easy word lists } && \text{(e.g., DC, Sp)}, \\
    G_2 &= \text{Metrics focusing on average sentence and word length} && \text{(e.g., FRES, FKGL, CLI)}, \\
    G_3 &= \text{Metrics accounting for syllables per word} && \text{(e.g., LW, Fog)}.
\end{aligned}
\]

Each formula $f \in G_i$ provides a grade-level $L_f$. The final grade-level calculation follows Algorithm~\ref{alg:readability}, which computes group-wise votes and determines the final grade through majority agreement or median value.

\begin{algorithm}[h!]
\caption{Grade-level Integration}
\label{alg:readability}
\KwIn{Grade-level estimates from \( G_1, G_2, G_3 \), where each formula \( f \) produces a grade-level \( L_f \).}
\KwOut{A final grade-level \( L_{\text{final}} \).}

\BlankLine
\ForEach(\tcp*[f]{Compute group-wise grade-level vote}){$G_i$ in \{\( G_1, G_2, G_3 \)\}}{
    Compute the group's grade-level vote \( L_{G_i} \) as:
    \[
    L_{G_i} = \bigcap_{f \in G_i} L_f.
    \]
    \eIf{\( L_{G_i} \neq \emptyset \)}{
        \(\mathrm{Vote}(G_i) \gets L_{G_i} \);
    }{
        \(\mathrm{Vote}(G_i) \gets \min \{ L_f \mid f \in G_i \} \);
    }
}

\BlankLine
\tcp{Determine final grade-level}
Let \( \{ L_{G_1}, L_{G_2}, L_{G_3} \} \) be the three computed votes\;

\BlankLine
\eIf{Two or more groups agree on the same grade-level}{
    \[
    L_{\text{final}} = \text{mode} \{ L_{G_1}, L_{G_2}, L_{G_3} \}.
    \]
}{
    \[
    L_{\text{final}} = \text{median} \left( \{ L_{G_1}, L_{G_2}, L_{G_3} \} \right).
    \]
}

\Return \( L_{\text{final}} \).
\end{algorithm}

\noindent The final result \( L_{\text{final}} \) belongs to the predefined set of grade levels:
\[
\{[1,2], [3,4], [5,6], [7,8,9], [10,11,12], [13+]\}.
\]

\subsection*{Model Training}
\label{subsec:model ft}
We perform supervised finetuning on GPT4o-mini via the OpenAI API to train six grade-specific models, spanning from lower elementary to adult (college+). We train each model on the corresponding subset of our grade-labeled question and answer corpus, classified based on our integrated readability metrics.  

\subsection*{Datasets}
\label{subsec:datasets}
We test our approach on real and synthetic datasets representing diverse grade levels. We use four datasets: ScienceQA ($\mathcal{D}_{SQ}$)~\cite{lu2022learn}, ELI5\_Category ($\mathcal{D}_{ELI}$)~\cite{eli5-category}, Natural Questions ($\mathcal{D}_{NQ}$)~\cite{47761}, and synthetic questions generated by GPT4o ($\mathcal{D}_{GPT}$).
\begin{itemize}
    \item{$\mathcal{D}_{SQ}$}: ScienceQA comprises multiple-choice questions across 21 educational domains for grades 1-12. We use this dataset to measure accuracy and convert the questions to an open-ended format using GPT4o for compatibility testing ($\mathcal{D}_{SQ2}$). We sample 10,876 and 10,427 questions for $\mathcal{D}_{SQ}$ and $\mathcal{D}_{SQ2}$ respectively.

    \item{$\mathcal{D}_{ELI}$}: ELI5\_Category contains questions from Reddit requiring explanatory multi-sentence answers. We sample 12,000 questions to evaluate compatibility for open-ended questions requiring detailed explanations.

    \item{$\mathcal{D}_{NQ}$}: Natural Questions contains real user questions submitted to Google search. We sample 24,000 questions to evaluate compatibility for naturally occurring questions.

    \item{$\mathcal{D}_{GPT}$}: We prompt GPT4o to create 740 questions across 54 subjects that require explanations and can be answered across all grade levels.
\end{itemize}



\section*{Acknowledgment}
This research was supported by the MSIT(Ministry of Science, ICT), Korea, under the Global Research Support Program in the Digital Field program)(RS-2024-00436680) supervised by the IITP(Institute for Information \& Communications Technology Planning \& Evaluation). This project is supported by Microsoft Research Asia. Jindong Wang was partially supported by The Commonwealth Cyber Initiative (CCI) program (H-2Q25-020), William \& Mary Faculty Research Award, and Modal Academic Compute Award. The authors acknowledge William \& Mary Research Computing for providing computational resources and/or technical support that have contributed to the results reported within this paper. The authors thank Xing Xie and Lexin Zhou for fruitful discussions.

\section*{Author Contributions}
J.O. and J.W. designed the main framework. J.O. proceeded with the experiments.  All authors reviewed and wrote the main manuscript text.

\section*{Competing Interests}
The authors declare no competing interests.

\clearpage
\bibliographystyle{naturemag}
\bibliography{refs}

\appendix
\section*{Supplementary Information}

\section{Related Work}

LLMs are getting huge interest as a tool for applications on various educational scenarios~\cite{wang2024largelanguagemodelseducation}. However, recent studies show that current LLMs fail to output grade-appropriate answers~\cite{gobara2024llmsimplicitlydeterminesuitable} regardless of how informative the prompt is~\cite{imperial-tayyar-madabushi-2023-flesch}. Current models consistently default to a high school~\cite{Hsu2024FreetextRG} or college+ level~\cite{rooein2023knowaudiencellmsadapt}, even when explicitly prompted to target a lower grade level. The root of this challenge lies in the training data itself, as foundational datasets such as Alpaca and Dolly exhibit a strong bias towards college+ level~\cite{10.1145/3627508.3638345}.

Beyond prompts, prior research has explored fine-tuning to align with different grade-levels. However, these efforts are confined to text simplification tasks that require a source text (e.g., summarization and paraphrasing)~\cite{malik-etal-2024-tarzan,ribeiro2023generating} and have shown poor performance for lower grades~\cite{tran2024readctrlpersonalizingtextgeneration}. Furthermore, the works have been limited to incorporating only a few educational metrics, given the difficulty of optimizing for multiple criteria simultaneously. In contrast, our work targets open-ended question answering, simulating a more realistic classroom setting of direct student-LLM interaction. By incorporating seven educational metrics, we achieve strong performance across all grade-levels,  proving effective even for the lower-elementary level. 

\section{Metrics}
\subsection{Educational Metrics}
\label{app:subsec_educational}
We compute text readability using multiple readability metrics. As each metric has different methods for mapping scores to grade levels, we standardize the results by classifying them into six predefined grade levels: 1 (lower elementary), 2 (middle elementary), 3 (high elementary), 4 (middle), 5 (high), and 6 (adult).
\subsubsection{Flesch Reading Ease}
The Flesch reading-ease score (FRES)~\cite{flesch1948new}, used by the U.S. Department of Defense, provides a numerical measure of text readability:
\[
  \text{FRES} = 206.835 
  \;-\; 1.015 \times \biggl(\frac{\text{total words}}{\text{total sentences}}\biggr) 
  \;-\; 84.6 \times \biggl(\frac{\text{total syllables}}{\text{total words}}\biggr).
\]
Higher FRES values correspond to text that is easier to read.

\subsubsection{Flesch–Kincaid grade level}
The Flesch--Kincaid Grade Level (FKGL)~\cite{kincaid1975derivation} formula is widely employed in education 
to estimate the U.S.\ grade level at which a text can be understood. 
It is calculated as:
\[
  \text{FKGL} 
  = 0.39 \times \biggl(\frac{\text{total words}}{\text{total sentences}}\biggr) 
  + 11.8 \times \biggl(\frac{\text{total syllables}}{\text{total words}}\biggr) 
  - 15.59.
\]
Higher FKGL scores correspond to texts requiring more advanced reading skills.
\subsubsection{Coleman--Liau Index}
The Coleman--Liau Index (CLI)~\cite{coleman1975computer} quantifies text readability by analyzing the average 
number of letters and sentences per 100 words. It is given by:
\[
  \text{CLI} 
  = 0.0588 \times L 
  \;-\; 0.296 \times S 
  \;-\; 15.8,
\]
where \(L\) is the average number of letters per 100 words 
and \(S\) is the average number of sentences per 100 words. 
Higher CLI scores correspond to texts requiring more advanced reading skills.

\subsubsection{Linsear Write}
Linsear Write (LW)~\cite{o1966gobbledygook} is a readability test developed for the U.S.\ Air Force 
to evaluate how easily technical manuals can be read. It is computed over a 
100-word passage as follows:

\begin{enumerate}
  \item Let $E$ be the number of ``easy'' words (words with at most two syllables),
    and $H$ be the number of ``hard'' words (three or more syllables).
  \item Let $S$ be the number of sentences in the 100-word sample.
  \item Compute the intermediate score $i$:
  \[
    i \;=\; \frac{\,E + 3H\,}{S}.
  \]
  \item Determine the final Linsear Write score ($\mathrm{LW}$):
  \[
    \mathrm{LW}
      = \begin{cases}
        \tfrac{i}{2}, & \text{if } i > 20, \\[6pt]
        \tfrac{i - 2}{2}, & \text{otherwise}.
        \end{cases}
  \]
\end{enumerate}
Higher LW scores correspond to texts requiring more advanced reading skills.

\subsubsection{Gunning--Fog Index}
The Gunning--Fog Index (\(\mathrm{Fog}\))~\cite{gunning1952technique} estimates the years of formal education required 
to comprehend a passage on the first reading. This index is defined as:
\[
  \mathrm{Fog} 
  \;=\; 0.4 \,\biggl(
  \tfrac{W}{S} \;+\; 100\,\tfrac{C}{W}\biggr).
\]
where \(W\) is the total number of words, \(S\) is the total number of sentences, and \(C\) is the number of ``complex'' words with three or more syllables. Higher \(\mathrm{Fog}\) values indicate text requiring more advanced reading skills.

\subsubsection{Dale--Chall Readability Formula}
The Dale--Chall readability formula (DC)~\cite{chall1995readability} assesses text complexity based on the proportion 
of unfamiliar words and sentence length, using a predefined list of 3000 words familiar to 80\% of fifth-grade students. It is defined as:
\[
\mathrm{DC}_{\text{raw}} = 0.1579 p + 0.0496 \dfrac{W}{S}.
\]
The final (adjusted) score is:
\[
\mathrm{DC} =
\begin{cases}
\mathrm{DC}_{\text{raw}} + 3.6365, & \text{if } p > 5, \\[8pt]
\mathrm{DC}_{\text{raw}}, & \text{otherwise}.
\end{cases}
\]
where \(D\) is the number of difficult words (words not in the predefined list), \(W\) is the total number of words, \(S\) is the total number of sentences, and \(p\) is the percentage of difficult words ($\dfrac{D}{W} \times 100$).
Higher \(\mathrm{DC}\) values indicate text requiring more advanced reading skills. 

\subsubsection{Spache Readability Formula}
The Spache readability formula ($\mathrm{Sp}$)~\cite{spache1953new} tests text readability for younger children. It uses a predefined list of approximately 1000 words and is defined as: 
\[
  \mathrm{Sp}
  \;=\;
  0.141\,S
  \;+\;
  0.086\,U
  \;+\;
  0.839.
\]
where \(S\) is the average sentence length (words per sentence) and  
and \(U\) is the percentage of unique unfamiliar words (words not in the predefined list). Higher $\mathrm{Sp}$ values indicate text requiring more advanced reading skills. 

\subsubsection{Automated Readability Index}
The Automated Readability Index (ARI)~\cite{smith1967automated} estimates the U.S.\ grade level required 
to comprehend a text. It is defined as:
\[
  \mathrm{ARI}
  =
  4.71\,\frac{C}{W}
  \;+\;
  0.5\,\frac{W}{S}
  \;-\;
  21.43.
\]
where \(C\) is the total number of characters (letters and digits), \(W\) is the total number of words, and \(S\) is the total number of sentences. Higher ARI scores indicate text requiring more advanced reading skills.

\subsection{Perplexity}
\label{app:subsec_ppl}
Perplexity (\(\mathrm{ppl}\)) predicts the probability of a token \(w_i\) given its preceding context \(w_1, \dots, w_{i-1}\). 
The perplexity is defined as:

\begin{equation*}
\mathrm{ppl} = \exp\Biggl(-\frac{1}{N}\sum_{i=1}^{N}\log p\bigl(w_i \mid w_1,\dots,w_{i-1}\bigr)\Biggr),
\label{eq:ppl}
\end{equation*}
where \(N\) is the number of tokens in the text dataset with $\log$ as the natural logarithm. 
Higher perplexity implies a broader probability spread across potential outcomes, indicating more diverse outputs. We use GPT2-XL for computing the probability.

\subsection{Diversity Gain}
\label{app:subsec_div}
Diversity gain~\cite{bilmes2022submodularitymachinelearningartificial} quantifies the extent to which a generated dataset brings diversity to the original (base) dataset.
The base dataset is defined as
\[
  D_{\text{base}} = \{\, x_i = (q_i,\; r_i,\; a_i) \}_{i=1}^N
\]
with \(N\) samples.
The new generated dataset is defined as
\[
  D_{\text{new}} = \{\, x_i = (q_i,\; r_i,\; a_i) \}_{i=1}^M
\]
with \(M\) samples.
The diversity gain of \(D_{\text{new}}\) relative to \(D_{\text{base}}\) is 
\[
  d_{\text{gain}} \;=\; \frac{1}{M} 
  \sum_{x_i \in D_{\text{new}}} \min_{x_j \in D_{\text{base}}} 
  \bigl\lVert f(x_i) \;-\; f(x_j) \bigr\rVert,
\]
where \(f\) is the feature extractor. We use $D_{\text{base}}$ as the collection of text outputs from the base model (GPT4o-mini) before finetuning and $D_{\text{new}}$ as the collection of text outputs from the grade-level finetuned models for all four datasets. We use \texttt{text-embedding-3-large} from OpenAI to extract the features.

\section{Extended Results}
\subsection{Fields and Subjects for Training Text Data}
\label{app:dataset generation}
We present the fields and subjects of questions used for finetuning based on the K-12 curriculum. 
\begin{figure*}[th]
    \centering
    \includegraphics[width=0.95\linewidth]{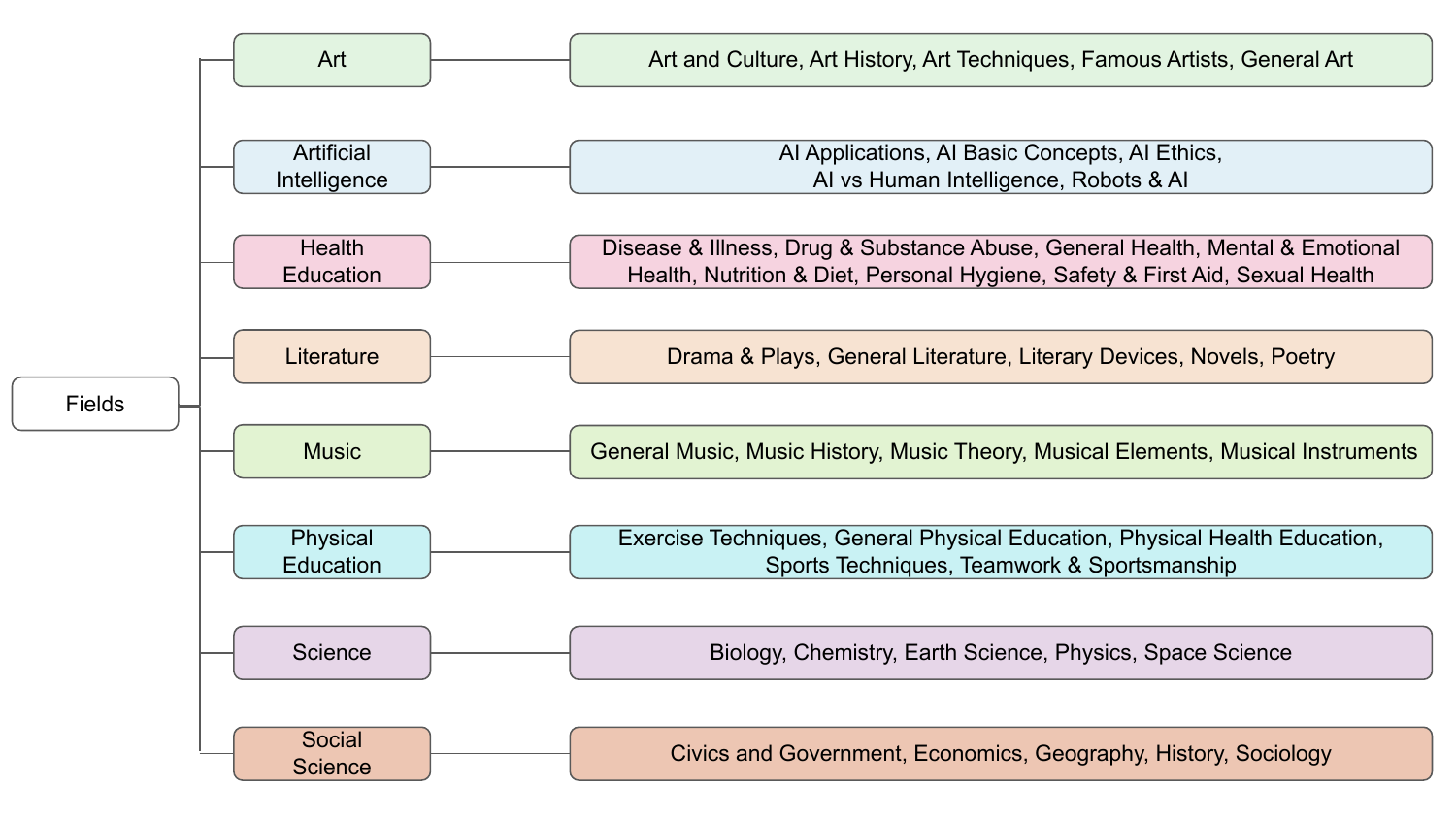}
    \caption{Fields and subjects of questions used for finetuning.} 
    \vspace{-0.3cm}
    \label{fig:subjects}
\end{figure*}
\clearpage
\subsection{Number of Q\&A for each Grade-level}

Table~\ref{tab:data_gen_dist} provides a detailed distribution of question \& answer pairs across the six grade-levels. We collect a total of 1,287,149 pairs. The amount of data for lower elementary is comparatively smaller than those for other grad-levels, due to the relatively strict thresholds set by the educational metrics. However, the absolute number of data remains substantial enough to facilitate effective finetuning. 
\begin{table*}[hbt]
\centering
\begin{tabular}{c@{\hspace{10pt}}|c@{\hspace{10pt}}|c@{\hspace{10pt}}|c@{\hspace{10pt}}|c@{\hspace{10pt}}|c@{\hspace{10pt}}|c@{\hspace{10pt}}}
\toprule
L-Elem&M-Elem&H-Elem&Middle&High&Adult&Total \\ 
\midrule
71,524&324,910&302,485&275,131&130,779&182,590&1,287,149\\

\bottomrule
\end{tabular}
\caption{Number of question \& answer pairs used for finetuning different models.}
\label{tab:data_gen_dist}
\end{table*}
\subsection{Extended Output Distribution}
\label{app:output_dist}
\begin{figure*}[h!]
    \centering
    \includegraphics[width=\linewidth]{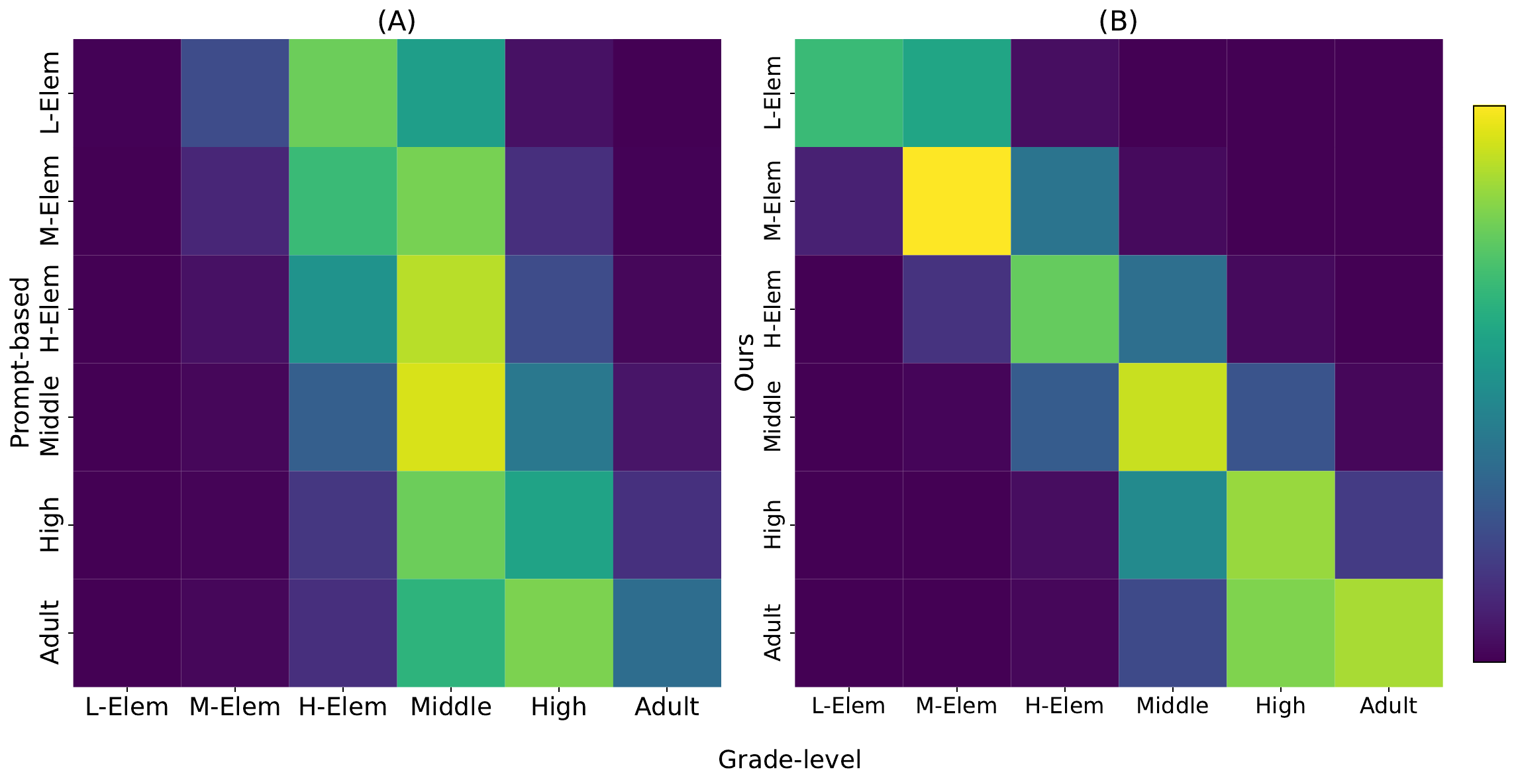}
    \caption{Comparison of output grade-level distributions from the prompt-based approach (A) and our method (B). The rows represent the targeted grade-level (prompt or model), and the columns show the classified grade-levels. Warmer colors (e.g., yellow) represent higher frequencies. For instance, the top row of (B) reveals the output classification results of the lower elementary model’s responses.}
    \vspace{-0.3cm}
    \label{fig:output_distribution}
\end{figure*}

We compare the output grade-level distributions from the prompt-based approach and ours in Figure~\ref{fig:output_distribution}. Overall, our approach shows a diagonal pattern, indicating stronger alignment with the intended grade levels than the prompt-based approach.

\clearpage
\subsection{Text Distribution in Data Generation for Different Prompts}
\label{app:instruction_prompt_dist}
We show the output grade-level distribution across different prompts, which are used for data generation. No single prompt consistently aligns with the intended grade level, highlighting the need for our proposed finetuning method.
\begin{figure*}[h!]
    \centering
    \includegraphics[width=0.95\linewidth]{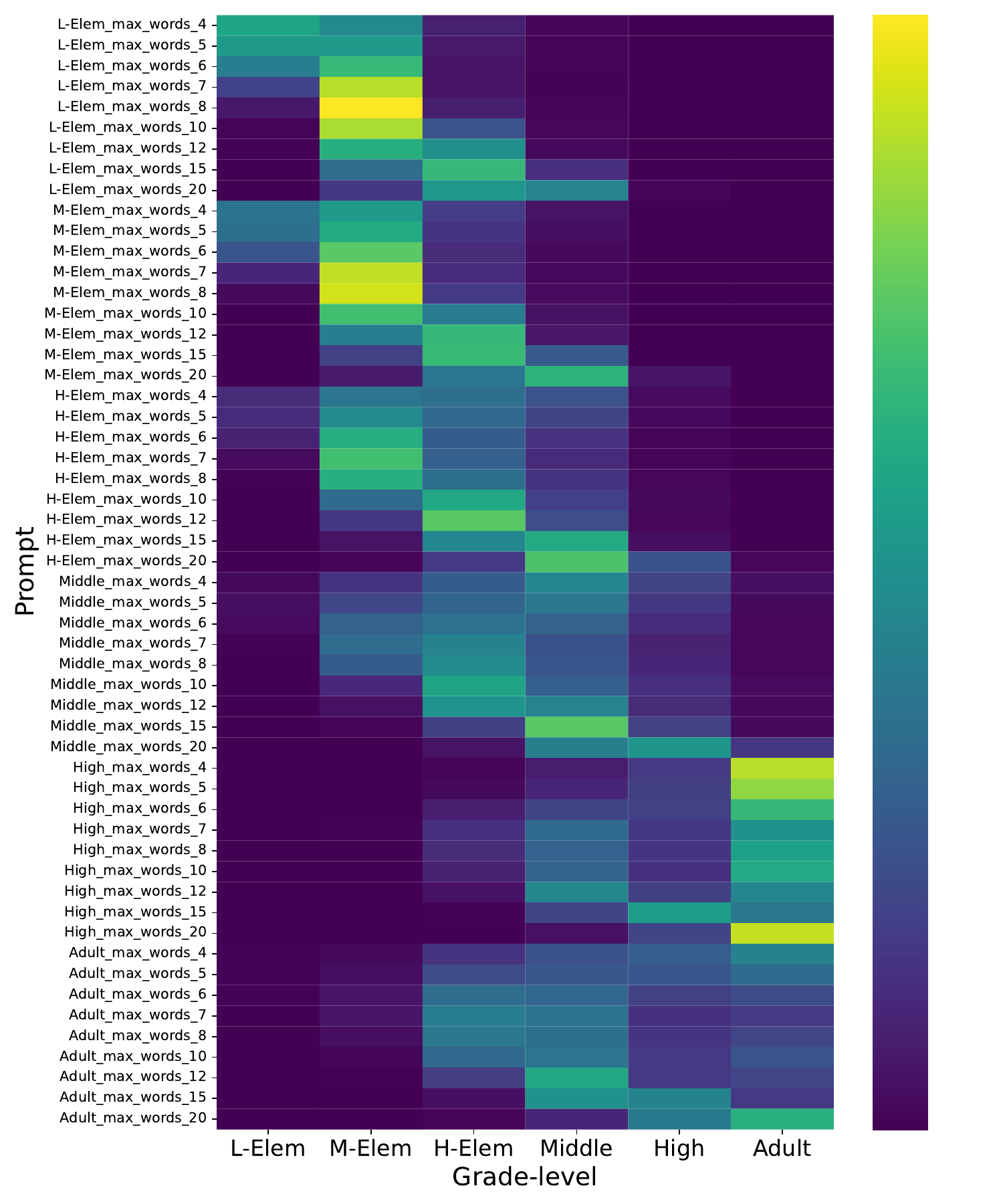}
    \caption{Grade distribution for different prompts in data generation. Each row represents a specific prompt, while the columns indicate the classified grade levels.} 
    \vspace{-0.3cm}
    \label{fig:prompt_distribution}
\end{figure*}

\clearpage
\subsection{Compatibility Results for Base LLMs}
In Table~\ref{tbl:base_results}, we compare the grade-level distribution of outputs among different base LLMs and their ARI scores. The results show that the current state-of-the-art LLMs tend to output with higher-grade readability, highlighting the challenge of generating content suitable for lower grade students. 
\begin{table}[h!]
\centering
\resizebox{0.85\textwidth}{!}{  
\begin{tabular}{c|c|c|c|c|c|c|c|c}
\toprule
Model & Dataset & L-Elem (1) & M-Elem (2) & H-Elem (3) & Middle (4) & High (5) & Adult (6) & ARI \\
\midrule
\multirow{4}{*}{GPT4o-mini}  
    & $\mathcal{D}_{GPT}$ & 0.00 & 0.00 & 7.57 & \bf44.59 & 34.73 & 13.11 & 5.25 \\
    & $\mathcal{D}_{SQ2}$ & 0.67 & 3.99 & 8.02 &31.53 & \bf43.81 & 14.74 & 4.21 \\
    & $\mathcal{D}_{ELI}$ & 0.00 & 0.08 & 1.59 & 37.93 & \bf49.26 & 11.14 & 5.24 \\
    & $\mathcal{D}_{NQ}$ & 0.03 & 1.14 & 11.00 & 35.69 & \bf36.78 & 15.37 & 4.82 \\
\midrule
\multirow{4}{*}{GPT4o}  
    & $\mathcal{D}_{GPT}$ & 0.00 & 0.27 & 7.16 & \bf41.35 & 35.27 & 15.95 & X \\
    & $\mathcal{D}_{SQ2}$ & 1.15 & 3.09 & 10.96 & 32.91 & \bf45.51 & 13.63 & X \\
    & $\mathcal{D}_{ELI}$ & 0.00 & 0.08 & 1.67 & 38.09 & \bf47.49 & 12.67 & X \\
    & $\mathcal{D}_{NQ}$ & 0.07 & 1.33 & 11.55 & \bf36.86 & 35.84 & 14.34 & X \\
\midrule
\multirow{4}{*}{LLaMA 3.3:70B}  
    & $\mathcal{D}_{GPT}$ & 0.00 & 0.68 & 11.89 & \bf46.35 & 32.16 & 8.92 & 5.06 \\
    & $\mathcal{D}_{SQ2}$ & 0.54 & 3.10 & 12.58 & 27.58 & \bf47.29 & 13.26 & 4.87 \\
    & $\mathcal{D}_{ELI}$ & 0.00 & 0.06 & 1.99 & 30.21 & \bf50.26 & 17.47 & 5.33 \\
    & $\mathcal{D}_{NQ}$ & 0.12 & 0.96 & 8.35 & 35.37 & \bf40.83 & 14.38 & 4.98 \\
\midrule
\multirow{4}{*}{LLaMA 3.1:70B}  
    & $\mathcal{D}_{GPT}$ & 0.00 & 0.14 & 13.38 & \bf51.35 & 28.51 & 6.62 & 5.04 \\
    & $\mathcal{D}_{SQ2}$ & 1.16 & 2.72 & 10.16 & 26.73 & \bf44.47 & 16.41 & 4.92 \\
    & $\mathcal{D}_{ELI}$ & 0.00 & 0.08 & 2.44 & 33.64 & \bf48.51 & 15.32 & 5.35 \\
    & $\mathcal{D}_{NQ}$ & 0.15 & 0.86 & 7.95 & 34.60 & \bf41.20 & 15.24 & 4.98 \\
\midrule
\multirow{4}{*}{LLaMA 3.1:8B}  
    & $\mathcal{D}_{GPT}$ & 0.00 & 0.41 & 10.95 & \bf48.92 & 32.16 & 7.57 & 5.09 \\
    & $\mathcal{D}_{SQ2}$ & 0.92 & 2.65 & 11.45 & 31.67 & \bf44.76 & 12.82 & 4.80 \\
    & $\mathcal{D}_{ELI}$ & 0.00 & 0.06 & 2.27 & 33.11 & \bf48.89 & 15.68 & 5.33 \\
    & $\mathcal{D}_{NQ}$ & 0.12 & 0.65 & 7.47 & 35.38 & \bf42.05 & 14.34 & 4.98 \\
\midrule
\multirow{4}{*}{Qwen 2.5:72B}  
    & $\mathcal{D}_{GPT}$ & 0.00 & 0.00 & 6.22 & 39.32 & \bf42.03 & 12.43 & 5.24 \\
    & $\mathcal{D}_{SQ2}$ & 0.25 & 3.30 & 10.44 & 31.98 & \bf49.78 & 9.60 & 4.85 \\
    & $\mathcal{D}_{ELI}$ & 0.00 & 0.03 & 1.30 & 33.83 & \bf53.59 & 11.25 & 5.28 \\
    & $\mathcal{D}_{NQ}$ & 0.01 & 0.27 & 5.76 & 33.67 & \bf44.16 & 16.13 & 5.07 \\
\midrule
\multirow{4}{*}{Gemma 2:27B}  
    & $\mathcal{D}_{GPT}$ & 0.00 & 0.41 & 8.24 & 29.73 & \bf37.43 & 24.19 & 5.29 \\
    & $\mathcal{D}_{SQ2}$ & 3.73 & 8.61 & 22.40 & 35.43 & \bf34.88 & 9.53 & 4.42 \\
    & $\mathcal{D}_{ELI}$ & 0.01 & 0.19 & 2.67 & 28.87 & \bf45.03 & 23.23 & 5.37 \\
    & $\mathcal{D}_{NQ}$ & 0.36 & 9.07 & 21.81 & \bf31.78 & 24.43 & 12.56 & 4.25 \\
\midrule
\multirow{4}{*}{Phi 4:14B}  
    & $\mathcal{D}_{GPT}$ & 0.00 & 0.00 & 4.87 & 36.89 & \bf42.16 & 13.24 & 5.44 \\
    & $\mathcal{D}_{SQ2}$ & 0.00 & 1.63 & 7.12 & 26.65 & \bf40.76 & 16.66 & 5.16 \\
    & $\mathcal{D}_{ELI}$ & 0.00 & 0.00 & 0.83 & 23.39 & \bf58.65 & 17.13 & 5.58 \\
    & $\mathcal{D}_{NQ}$ & 0.03 & 0.70 & 8.56 & 30.44 & \bf39.13 & 21.15 & 5.04 \\
\midrule
\multirow{4}{*}{Mixtral 8x7B}  
    & $\mathcal{D}_{GPT}$ & 0.00 & 1.36 & 15.20 & \bf37.04 & 29.44 & 16.96 & 5.07 \\
    & $\mathcal{D}_{SQ2}$ & 2.33 & 2.95 & 14.40 & 36.30 & \bf44.15 & 9.79 & 4.72 \\
    & $\mathcal{D}_{ELI}$ & 0.01 & 0.16 & 4.60 & 31.31 & \bf44.97 & 18.95 & 5.24 \\
    & $\mathcal{D}_{NQ}$ & 0.14 & 0.93 & 10.39 & 36.81 & \bf39.22 & 12.52 & 4.88 \\
\bottomrule
\end{tabular}
} 
\caption{Target \% and ARI values for various base LLMs across four datasets. Each column (lower elementary to adult) shows the proportion of outputs classified at that grade level, while the final column presents the ARI value. Bolded entries indicate the highest proportion for each model and dataset pair.}
\label{tbl:base_results}
\end{table}
\clearpage

\subsection{Compatibility Results for Prior Approaches and Ours}
\label{app:res_compat_detail}
We provide a comprehensive breakdown of grade-level output distribution for several models across four datasets, comparing prompt-based approach and ours. Table~\ref{tbl:extended_compat} includes both the percentage of outputs correctly aligned with each target grade level (Target \%) and the ARI values. Notably, our finetuned models often achieve stronger alignment at the targeted grade-levels.

\begin{table}[h!]
\centering
\resizebox{0.95\textwidth}{!}{  
\begin{tabular}{c|c|c|cc|cc|cc|cc|cc|cc}
\toprule
        &  &  & \multicolumn{2}{c|}{L-Elem (1)} & \multicolumn{2}{c|}{M-Elem (2)} & \multicolumn{2}{c|}{H-Elem (3)} & \multicolumn{2}{c|}{Middle (4)} & \multicolumn{2}{c|}{High (5)} & \multicolumn{2}{c}{Adult (6)} \\
\midrule
        & Model & Dataset & Target \% & ARI & Target \%& ARI & Target \%& ARI & Target \%& ARI & Target \%& ARI & Target \%& ARI \\
\midrule

\multirow{4}{*}{\shortstack{Prompt \\Based}} 
        &\multirow{4}{*}{\shortstack{GPT4o \\ mini\\(Baseline)}}&\small $\mathcal{D}_{GPT}$ & 0.13 & 3.35 & 11.22 & 3.77 & 36.22 & 3.99 & 50.68 & 4.36 & 27.30 & 4.87 & 22.16 & 5.40 \\
        &&\small $\mathcal{D}_{SQ2}$ & 8.04 & 2.67 & 16.80 & 3.40 & 37.28 & 3.84 & 54.30 & 4.48 & 47.32 &  5.10 & 21.95 & 5.13 \\
        &&\small $\mathcal{D}_{ELI}$ & 0.05 & 3.51 & 5.36 & 3.77 & 37.99 & 3.96 & 64.18 &  4.32 & 32.58 &  4.82 & 14.96 & 5.35 \\
        &&\small $\mathcal{D}_{NQ}$ & 0.25 & 3.29 & 5.38 & 3.75 & 25.42 & 4.07 & 49.94 & 4.40 & 35.55 & 4.66 & 18.00 & 4.90 \\
    \cline{2-15}
        & \multirow{4}{*}{\shortstack{GPT4o}} & \small $\mathcal{D}_{GPT}$ & 0.27 & 3.39 & 14.46 & 3.61 & 31.62 & 3.89 & 52.70 & \bf 4.26 & 35.14 & \bf 4.96 & 21.62 & 5.31 \\
        &  & \small $\mathcal{D}_{SQ2}$ & 4.70 & 2.98 & 14.17 & 3.60 & 32.33 & 3.95 & 52.39 & 4.51 & 48.74 & 5.14 & 16.95 & 5.04 \\
        &  & \small $\mathcal{D}_{ELI}$ & 0.13 & 3.49 & 8.71 & 3.63 & 41.29 & 3.83 & 62.38 & 4.21 & 26.47 & 4.60 & 13.06 & 5.22\\
        &  & \small $\mathcal{D}_{NQ}$ & 1.20 & 2.97 & 14.31 & 3.36 & 34.09 & 3.78 & 56.67 & \bf 4.22 & 35.34 & 4.67 & 15.53 & 4.92 \\
\cline{2-15}
    & \multirow{4}{*}{\shortstack{LLaMA \\3.3:70B}} & \small $\mathcal{D}_{GPT}$ & 0.14 & 3.22 & 10.81 & 3.65 & 31.89 & 4.04 & 48.51 & 4.44 & 29.19 & 4.81 & 12.57 & 5.19 \\
    &  & \small $\mathcal{D}_{SQ2}$ & 6.10 & 2.48 & 17.69 & 3.38 & 34.50 & 3.89 & 47.81 & 4.68 & 49.86 & 5.16 & 14.66 & 5.06 \\
    &  & \small $\mathcal{D}_{ELI}$ & 0.02 & 3.54 & 2.56 & 3.92 & 18.30 & 4.29 & 53.89 & 4.73 & 47.74 & 5.11 & 18.87 & 5.35 \\
    &  & \small $\mathcal{D}_{NQ}$ & 0.59 & 3.31 & 6.04 & 3.86 & 17.73 & 4.34 & 44.25 & 4.72 & 41.80 & 4.93 & 14.53 & 5.08 \\
\cline{2-15}
    & \multirow{4}{*}{\shortstack{LLaMA \\3.1:70B}} & \small $\mathcal{D}_{GPT}$ & 0.27 &3.20& 10.68 & 3.69 & 34.60 & 4.03 & 53.51 & 4.39 & 24.73 & 4.75 & 9.19 & 5.10  \\
    &  & \small $\mathcal{D}_{SQ2}$ & 5.37 & 2.49 & 18.98 & 3.35 & 31.09 & 3.96 & 49.43 & 4.65 & 46.57 & 4.96 & 11.74 & 4.97 \\
    &  & \small $\mathcal{D}_{ELI}$ & 0.03 & 3.52 & 2.76 & 3.87 & 19.89 & 4.29 & 56.10 & 4.70 & 43.50 & \bf 5.06 & 13.99 & 5.30 \\
    &  & \small $\mathcal{D}_{NQ}$ & 0.66 & 3.28 & 5.40 & 3.90 & 16.32 & 4.40 & 45.08 & 4.67 & 39.04 & 4.84 & 15.45 & 5.07 \\
\cline{2-15}
    & \multirow{4}{*}{\shortstack{LLaMA \\3.1:8B}} & \small $\mathcal{D}_{GPT}$ & 0.14 & 3.26 & 13.67 & 3.58 & 40.96 & 3.92 & 46.22 & 4.35 & 27.70 & 4.83 & 10.95 & 5.18 \\
    &  & \small $\mathcal{D}_{SQ2}$ & 6.80 & 2.50 & 23.76 & 3.17 & 38.69 & 3.78 & 49.52 & 4.48 & 44.19 & 4.91 & 10.21 & 4.85 \\
    &  & \small $\mathcal{D}_{ELI}$ & 0.02 & 3.49 & 4.10 & 3.77 & 24.85 & 4.16 & 57.44 & 4.64 & 43.17 & 5.04 & 12.85 & 5.27 \\
    &  & \small $\mathcal{D}_{NQ}$ & 0.49 & 3.32 & 6.10 & 3.82 & 17.98 & 4.31 & 45.94 & 4.61 & 39.77 & 4.86 & 14.12 & 5.06 \\
\cline{2-15}
    & \multirow{4}{*}{\shortstack{Qwen\\2.5:72B}} & \small $\mathcal{D}_{GPT}$ & 0.00 & 3.59 & 9.87 & 3.70 & 42.97 & 3.80 & 63.38 & 4.43 & 37.97 & 5.07 & 13.24 & 5.27 \\
    &  & \small $\mathcal{D}_{SQ2}$ & 4.16 & 2.92 & 19.10 & 3.33 & 45.65 & 3.62 & \bf 59.90 & 4.42 & 51.51 & \bf 5.03 & 9.81 & 4.86 \\
    &  & \small $\mathcal{D}_{ELI}$ & 0.03 & 3.57 & 5.19 & 3.68 & 40.64 & 3.87 & 68.9 & 4.44 & 49.29 & 5.10 & 8.88 & 5.22 \\
    &  & \small $\mathcal{D}_{NQ}$ & 0.10 & 3.52 & 5.07 & 3.80 & 20.16 & 4.28 & 47.16 & 4.67 & 43.34 & 4.97 & 15.57 & 5.03 \\
\cline{2-15}
    & \multirow{4}{*}{\shortstack{Gemma\\2:27B}} & \small $\mathcal{D}_{GPT}$ & 1.14 & 2.97 & 17.16 & 3.36 & 41.76 & 3.64 & 54.73 & 3.94 & 25.14 & 4.55 & 29.46 & 5.38 \\
    &  & \small $\mathcal{D}_{SQ2}$ & 15.53 & 1.92 & 37.91 & 2.91 & 47.49 & \bf3.28 & 46.32 & \bf 3.74 & 28.12 & 4.47 & 12.65 & 4.65 \\
    &  & \small $\mathcal{D}_{ELI}$ & 0.13 & 3.19 & 11.99 & 3.53 & 43.18 & 3.76 & 59.49 & \bf 4.11 & 27.61 & 4.71 & 25.50 & 5.42 \\
    &  & \small $\mathcal{D}_{NQ}$ & 7.58 & 2.34 & 31.97 & 2.77 & 40.49 & \bf3.08 & 40.26 & 3.58 & 21.74 & 4.02 & 12.70 & 4.22 \\
\cline{2-15}
    & \multirow{4}{*}{\shortstack{Phi\\4:14B}} & \small $\mathcal{D}_{GPT}$ & 0.00 & 3.72 & 9.87 & 3.86 & 39.46 & 4.02 & 55.27 & 4.35 & 31.49 & 5.07 & 12.57 & \bf 5.43 \\
    &  & \small $\mathcal{D}_{SQ2}$ & 1.71 & 3.05 & 16.12 & 3.49 & 39.95 & 3.81 & 58.25 & 4.52 & 48.29 & 5.20 & 12.65 & 5.07 \\
    &  & \small $\mathcal{D}_{ELI}$ & 0.00 & 3.84 & 3.48 & 3.93 & 35.27 & 4.05 & \bf 67.16 & 4.47 & 46.45 & 5.18 & 12.56 & 5.51 \\
    &  & \small $\mathcal{D}_{NQ}$ & 0.10 & 3.72 & 3.47 & 4.05 & 17.51 & 4.32 & 45.15 & 4.77 & 47.91 & 5.20 & 23.06 & 5.17 \\
\cline{2-15}
    & \multirow{4}{*}{\shortstack{Mixtral\\:8x7B}} & \small $\mathcal{D}_{GPT}$ & 0.68 & 3.83 & 7.57 & 3.97 & 18.02 & 4.62 & 50.81 & 4.54 & 34.39 & 4.92 & 21.94 & 5.22 \\
    &  & \small $\mathcal{D}_{SQ2}$ & 1.96 & 3.16 & 10.60 & 3.87 & 15.96 & 4.55 & 45.54 & 4.61 & 45.73 & \bf 5.03 & 11.17 & 4.75 \\
    &  & \small $\mathcal{D}_{ELI}$ & 0.02 & 4.28 & 2.00 & 4.41 & 4.81 & 5.10 & 44.56 & 4.93 & 48.85 & 5.25 & 20.65 & 5.35 \\
    &  & \small$\mathcal{D}_{NQ}$ & 0.09 & 3.96 & 2.16 & 4.46 & 7.01 & 4.94 & 41.49 & 4.81 & 44.66 & 5.04 & 13.32 & 4.95 \\
\midrule
\multirow{4}{*}{Ours} 
       & \multirow{4}{*}{\shortstack{Finetuned\\GPT4o \\ mini}}&\small $\mathcal{D}_{GPT}$ & \bf67.30 & \bf 1.32 & \bf82.97 & \bf 2.14 & \bf73.11 & \bf 3.42 & \bf70.54 & 4.56 & \bf63.24 & 5.33 & \bf73.78 &  5.40 \\
       & &\small $\mathcal{D}_{SQ2}$ & \bf65.99 & \bf 1.21 & \bf67.08 & \bf 2.20 & \bf52.74 &  3.56 & 58.83 & 4.45 & \bf65.40 & 5.26 & \bf 43.16 & \bf 5.67 \\
       & &\small $\mathcal{D}_{ELI}$ & \bf57.41 & \bf 1.36 & \bf72.82 & \bf 2.24 & \bf60.00 & \bf 3.55 & 65.18 & 4.38 & \bf64.68 & 5.34 & \bf66.16 & \bf 5.88 \\
       & &\small $\mathcal{D}_{NQ}$ & \bf46.65 & \bf 1.37 & \bf61.90 & \bf 2.07 & \bf58.21 &  3.27 & \bf58.71 & 4.23 & \bf49.07 & \bf 4.99 & \bf33.54 & \bf 5.49 \\

\bottomrule
\end{tabular}
} 
\caption{Extended performance comparison across multiple models, grade levels, and datasets. Each cell displays the Target\% and the corresponding ARI value. Bold values indicate the highest performance for a particular dataset–grade combination.}
\label{tbl:extended_compat}

\end{table}
\clearpage

\subsection{Survey Results with GPT}
\label{app:survey_gpt}
\begin{figure*}[ht]
    \centering
    \includegraphics[width=0.4\linewidth]{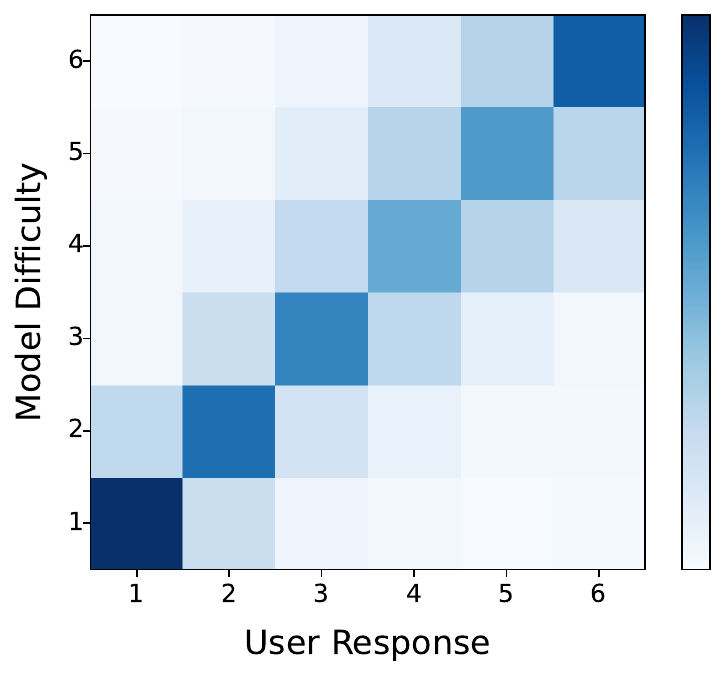}
    \caption{Survey results for type 1 questions by GPT4o.} 
    \vspace{-0.3cm}
    \label{fig:corr_gpt}
\end{figure*}

\begin{figure*}[ht]
    \centering
    \includegraphics[width=0.8\linewidth]{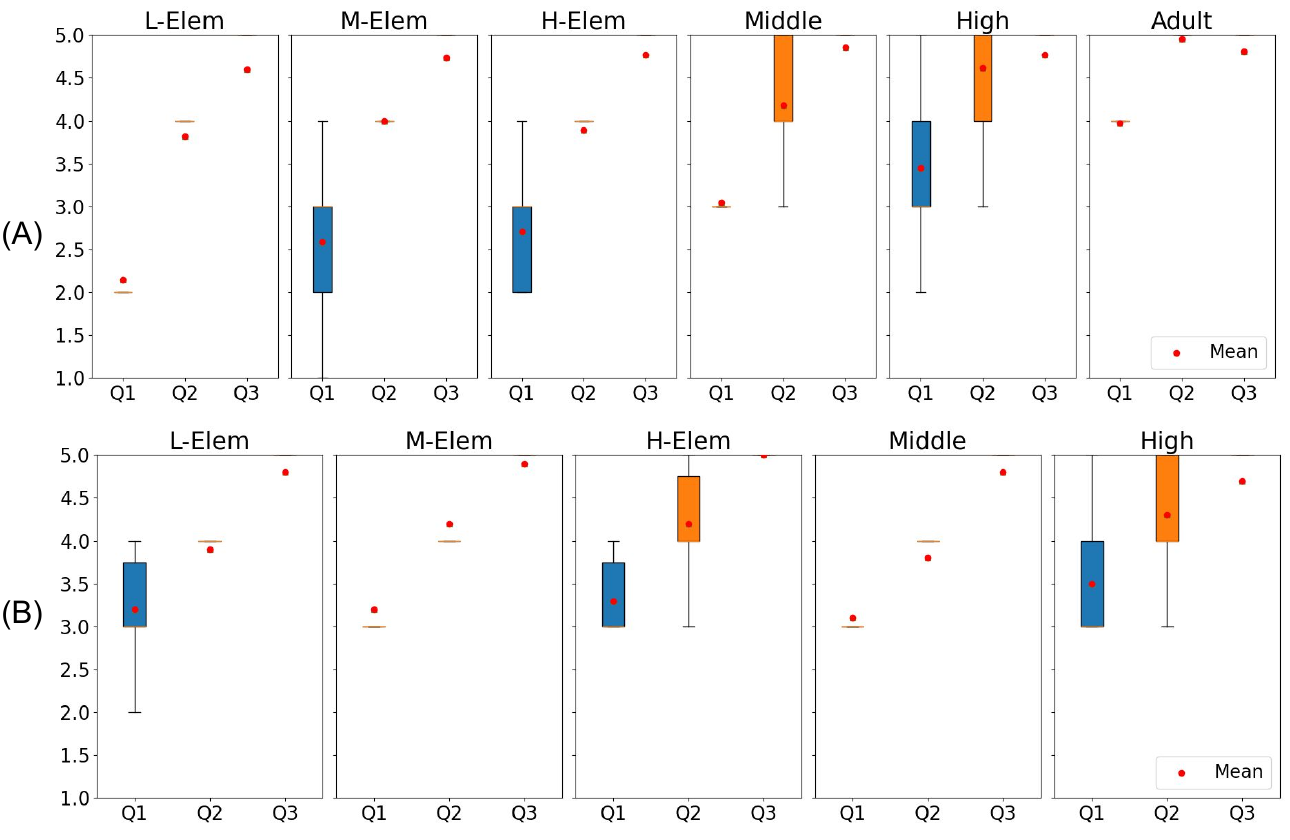}
    \caption{Survey results for type 2 questions by GPT4o.} 
    \vspace{-0.3cm}
    \label{fig:survey_t2_gpt}
\end{figure*}

We present the results on the survey results from GPT4o. We feed GPT4o with the same set of survey questions posed to human participants. For type 1 questions, we observe Kendall's $\tau$ coefficient as 0.72 and an L1 distance of [0.419, 0.585, 0.635, 0.831, 0.724, 0.601], along with a strong alignment between model outputs and GPT4o's perceptions (see Figure~\ref{fig:corr_gpt}). For type 2 questions, answer comprehensibility and accuracy consistently remains high regardless of the question difficulty (see Figure~\ref{fig:survey_t2_gpt}). These high correlation between GPT4o’s output ratings and the ratings provided by human participants robustly validates our framework's effectiveness. 
\subsubsection{Type 1 Question}
\subsubsection{Type 2 Question}
\section{Exploring Worldview of Models--Output Distribution}
\label{app:sec_worldview}

We analyze the different worldviews of models through output analysis across the experimental datasets. Since educational metrics primarily capture sentence and word-level complexity, we present three analyses of these textual aspects.

\subsection{Sentence Length Distribution}
\begin{figure*}[ht]
    \centering
    \includegraphics[width=0.7\linewidth]{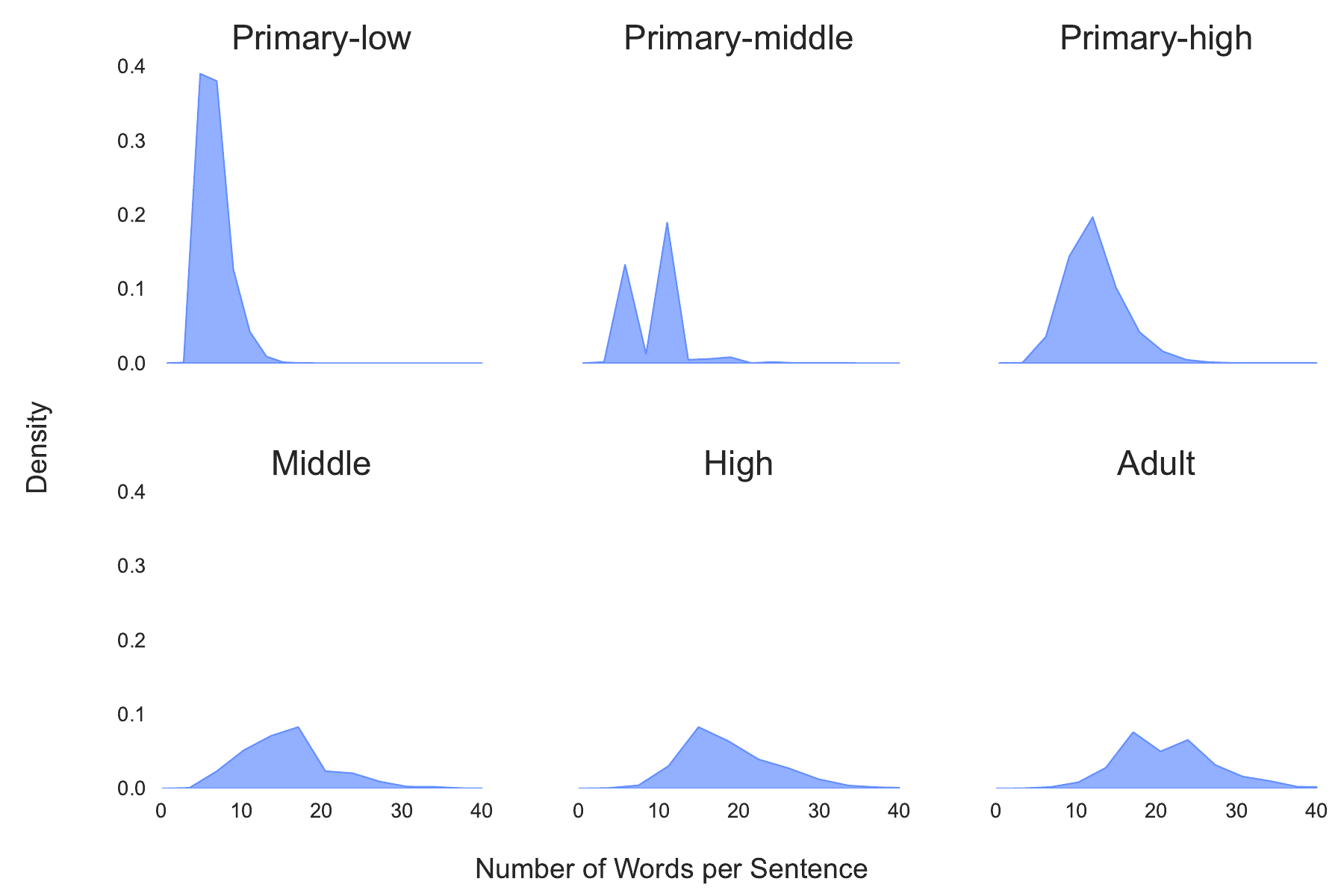}
    \caption{Sentence length distribution across different models.} 
    \vspace{-0.3cm}
    \label{fig:sent_dist}
\end{figure*}

As shown in Figure~\ref{fig:sent_dist}, sentence length increases for higher-grade models. These models produce more detailed explanations that require advanced reading skills.
Lower grade models generate shorter statements that match younger learners' language needs \cite{balthazar2024sentences}. Each model's worldview shifts with its target grade level, from concise simplicity to detailed explanation.

\subsection{Word Difficulty}
\begin{figure*}[ht]
    \centering
    \includegraphics[width=0.9\linewidth]{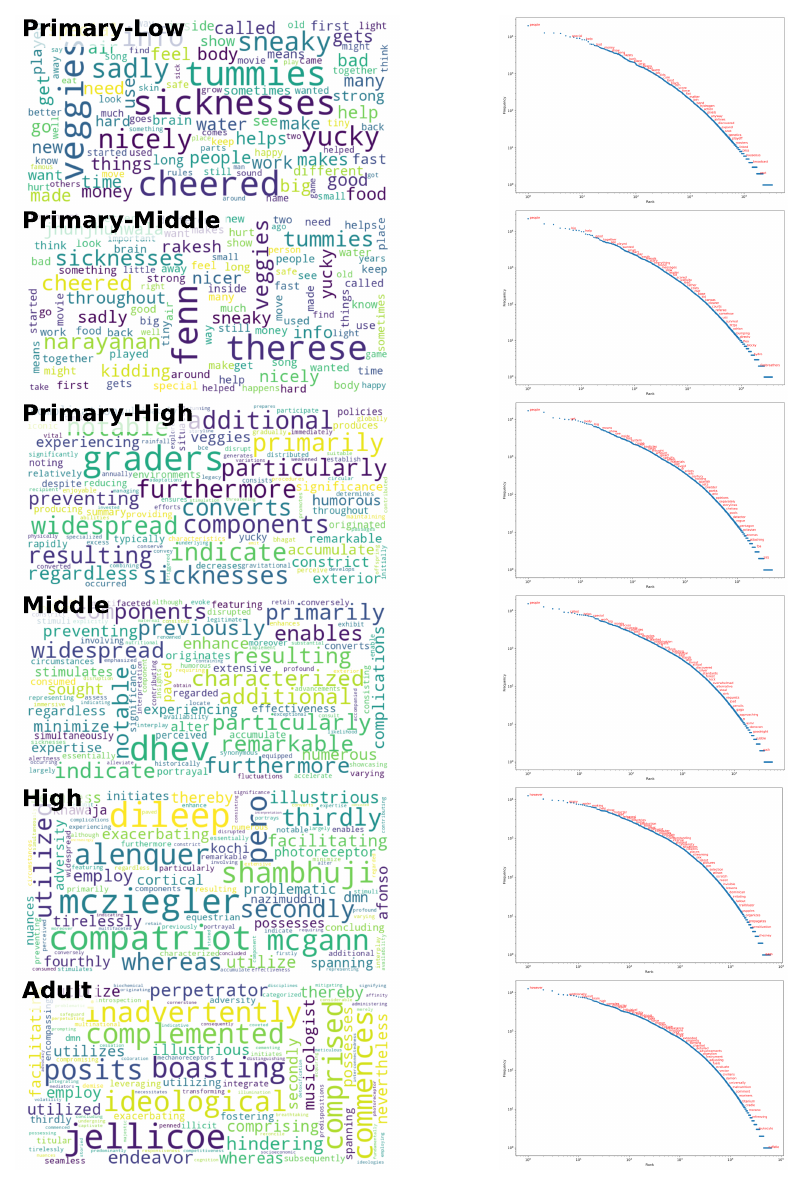}
    \caption{Word Cloud and Zipfian Distribution} 
    \vspace{-0.3cm}
    \label{fig:worddist}
\end{figure*}

A key component of model worldview analysis involves examining output vocabulary. Humans use different words as they develop, even when describing identical situations. We visualize outputs through word clouds and Zipfian distributions~\cite{zipf2013psycho}, applying BM25~\cite{robertson1995okapi} to highlight distinctive terms while minimizing common words like ``people'' and ``like.'' We exclude words appearing fewer than 30 times to eliminate outliers. Figure~\ref{fig:worddist} reveals vocabulary shifts across different models, with higher-grade models using more complex terms that reflect advanced linguistic development.

\clearpage
\section{Survey Details}
\label{app:sec_survey}
We provide further details on how we distribute and design the surveys.
\subsection{Survey Distribution Details}

For Survey 1, we sampled questions from $\mathcal{D}_{NQ}$, consisting of real user questions submitted to Google Search. We assessed whether finetuned model answers to natural questions were both comprehensible for target grades and factually accurate. We provided 5 Type 1 questions and 18 Type 2 questions to each participant with compensation of 7000 Korean Won (approximately 5 USD) for an estimated 30 minutes of work. We recruited 108 English-speaking participants from the Republic of Korea, the USA, and England. To ensure question diversity, we created 18 distinct forms, each completed by six participants. 

For Survey 2, we sampled questions from $\mathcal{D}_{SQ2}$ to measure the comprehensibility and accuracy of finetuned models for grade-appropriate questions. $\mathcal{D}_{SQ2}$ contains questions classified for grades 1-12 based on difficulty. We provided 10 Type 2 questions per grade level to 100 participants from Amazon Mechanical Turk. Participants received 1.5 USD for an estimated 8 minutes of work. 

\subsection{Type 1 Question}
\begin{itemize}
\item Q: Assign each answer to a unique grade level based on how comprehensible/understandable it would be to students of that grade.
\end{itemize}

\subsection{Type 2 Questions}
\begin{itemize}
        \item Q1: Is the question answerable by X-grade students (regardless of the given answer)?
            \begin{itemize}
                \item 5: Any X-grade student can answer this question.
                \item 4: Most X-grade students can answer this question.
                \item 3: It is likely that X-grade students can answer this question.
                \item 2: Most X-grade students cannot answer this question.
                \item 1: It is impossible for X-grade students to answer this question.
            \end{itemize}
        \item Q2: Is the answer comprehensible by X-grade students (regardless of the given question)?
            \begin{itemize}
                \item 5: Any X-grade student can comprehend this answer.
                \item 4: Most X-grade students can comprehend this answer.
                \item 3: It is likely that X-grade students can comprehend this answer. 
                \item 2: Most X-grade students cannot comprehend this answer.
                \item 1: It is impossible for X-grade students to comprehend this answer.
            \end{itemize}
        \item Q3: Is the answer consistently answering the question?
            \begin{itemize}
                \item 5: The answer answers the corresponding question.
                \item 4: The answer mostly follows the corresponding question.
                \item 3: The answer roughly follows the corresponding question.
                \item 2: The answer hardly follows the corresponding question.
                \item 1: The answer is completely irrelevant to the corresponding question.
            \end{itemize}
    \end{itemize}

\end{document}